\documentstyle[aps,eqsecnum,preprint]{revtex}
\begin{document}
\preprint{
\rightline{\vbox{\hbox{\rightline{MSUNSCL-1050}} 
\hbox{\rightline{LU-96-01}}}}
         }
\title{Low-mass dileptons at the SPS}

\author{J. Murray and W. Bauer}
\address{
National Superconducting Cyclotron Laboratory, Michigan State University, 
East Lansing, MI 48824}

\author{K. Haglin}
\address{
Department of Physics, Lawrence University, Appleton, WI 54912}
\maketitle

\begin{abstract}

We use a simple QCD-based model to study particle production in 
S+Au collisions at 200 GeV/n.  A requisite consistency is met
for the hadronic observables ($\pi^0$ and $\pi^-$ spectra) while pursuing
estimates for $e^+e^-$ production.  Since
radiative decays of initially produced hadrons has accounted for
only a portion of the observed dileptons at the CERN SPS,
we search for additional mechanisms. By including contributions from 
prompt secondary hadronic scatterings, $\pi \rho \to \pi e^+
e^-$, and adding to $\pi^+\pi^-$ annihilation and hadronic decays, 
the ``excess'' dilepton signal can possibly be interpreted.

\end{abstract}

\narrowtext

\section{Introduction}

Measuring and analyzing electromagnetic radiation from heavy-ion 
collisions represents a significant experimental challenge compared to 
hadronic signals owing to the relatively small cross sections.  The
additional information they provide certainly justifies the undertaking.  
Hadrons produced in the initial stages of the collision interact 
on average several times before leaving the reaction zone.  
Consequently, any information embedded in hadronic dynamics is completely 
masked by multiple scatterings. Dileptons are not disturbed by the
hadronic environment even though they are produced at all stages of the
collisions as they have long mean free paths.  They are dubbed
``clean probes'' of the collision dynamics. 

Recent results from CERN\cite{ceres} have brought about a surge of activity
in search of quantitative interpretation.  The proton-induced reactions
(p+Be and p+Au at 450 GeV) are consistent with 
predictions from  primary particle production and subsequent 
radiative and/or Dalitz decays suggesting that the
$e^+e^-$ yields are fairly well understood.  Yet, the heavy-ion data (S+Au
at 200 GeV/n) show a significant excess as compared to the same model for 
meson production and electromagnetic decays.  When integrated over pair
invariant mass up to 1.5 GeV, the number of electron pairs exceeded
the ``cocktail'' prediction by a factor of 5$\pm 2$.  It is clear that
two-pion annihilation contributes in the heavy-ion reactions
as fireball-like features emerge and support copious pion production.
Vector dominance arguments would naturally lead to extra production
around the rho mass.  Yet, the excess is most pronounced between the
two-pion threshold and the rho mass.

The nature of the enhancement suggests several possibilities. Medium
modifications resulting in a shifted rho mass could be
responsible\cite{rhoshift}. Along these lines, effects arising
from a modified pion dispersion relation have been investigated
\cite{pidisp}. Kapusta, Kharzeev, and McLerran  suggest that the enhancement 
may be due to a shift in the $\eta^{\prime}$ mass. If medium properties 
are favorable, the $\eta^{\prime}$ can be considered a Goldstone boson 
with a small mass which implies
that the cross section for producing an $\eta^{\prime}$ is larger. This in 
turn could boost the number of dileptons in the region of 
observed excess\cite{goldstone}. 

Secondary scattering of pions and other
resonances has also been studied\cite{haglin} focusing on the
role of the $a_{1}$ through $\pi \rho\to a_{1}\to \pi\ e^+e^-$.
The contribution was shown to be relevant but not sufficient for
interpreting the data.
We extend the secondary scattering investigation in the present
calculation by including non-resonance dilepton-producing 
$\pi\rho\to\pi\ e^+e^-$ reactions\cite{first_suggestion}.  We shall 
organize our paper in the
following way.  In Sec.~\ref{sec:primary} we discuss the event
generator providing the basis of the reaction description and we
compare to pion spectra from experiment.  Then in Sec.~\ref{sec:secondary}
we introduce an algorithm for estimating secondary scattering.
The prompt, or non-resonance, $\pi\rho\to\pi\ e^+e^-$ reactions are
modeled with an effective field theory.  A brief description of
gauge-invariantly introducing strong-interaction form factors also 
appears.  In Sec.~\ref{sec:norm} we discuss normalization and 
acceptance effects, followed by the results and conclusions in
Sec.~\ref{sec:results}.

\section{Primary Scattering} 
\label{sec:primary}
Future collider energies, several thousand GeV per nucleon in the 
center of mass, probe distances much smaller than the nucleon.
Models must of course incorporate QCD to describe the subnucleonic features.
Quarks and gluons then comprise the 
appropriate degrees of freedom for a QCD transport theory.  They
are propagated through spacetime approximating 
the dynamics of collisions to be explored at RHIC and LHC
\cite{geiger1,geiger2,kort}.  Evolution continues until soft processes 
dominate and hadronization occurs.  

Whether or not a quark-gluon plasma can be experimentally detected depends
largely on the characteristics of the collision in its absence, something
we shall call background.  In order to better quantify this background,
simulations without ``built-in'' plasma formation that still assume a QCD
description of nucleon scattering must be explored.
HIJING, developed by Wang and Gyulassy\cite{hijing}, is precisely this
type of model and has been used to look at multiple minijet production,
shadowing and jet quenching in pA and AA collisions.

The simulation we develop is similar to HIJING. It is based on a simple
prescription that uses QCD to characterize the individual nucleon-nucleon
collisions and uses Glauber-type geometry to determine the scaling. 
The kinematics of the nucleon-nucleon collisions are handled by PYTHIA
and JETSET\cite{sjostrand}, high energy event-generators using 
QCD matrix elements as well as the Lund fragmentation scheme. A somewhat 
detailed description of the model is outlined below.

\begin{itemize}
\item[$\bullet$] {\em Initialization of nucleons inside nuclei:} The nucleons
are
positioned randomly inside each nucleus according to the size of the nucleus
and are given random Fermi momentum in the $x$-$y$ plane and are
given $z$-momentum proportional to the lab energy (or center of mass energy).

\item[$\bullet$] {\em Number of collisions:} The number of collisions is
determined geometrically \cite{ncoll}. For a proton-nucleus collision, 

\begin{equation}
n(b)={\sigma}_{NN} \int dx\, dy\,
dz \, \rho (\sqrt{b^2+z^2})
\end{equation}

\narrowtext
For nucleus-nucleus collisions,

\widetext
\begin{equation}
N(b)={\sigma}_{NN} \int dx \, dy \, dz_{1}\,dz_{2}\,
\rho_{A}(\sqrt{x^2+y^2+{z_1}^2}) \, \rho_{B}(\sqrt{x^2+{(y-b)}^2
+{z_2}^2})
\end{equation}

\item[$\bullet$] {\em Picking scattering partners:} Two nucleons are 
chosen at random from
each nuclei and are allowed to scatter when, and if, they meet several 
criteria: First, the two nucleons cannot have scattered previously. Second,
the nucleons must be within one cross-sectional radius of one
another in the transverse beam direction, 

\narrowtext
\begin{equation}
\sqrt{(x_t-x_p)^2+(y_t-y_p)^2} \leq \sqrt{\sigma_{NN}/{\pi}}
\end{equation}

Thirdly, the pair must be moving toward one another in the transverse plane. 
And lastly, the center of mass energy, $\sqrt{s}$, must be above 6 GeV. This
limit is chosen because it is on the order of the energy where
perturbative QCD is no longer applicable.
If the two nucleons meet these criteria, they are allowed to
scatter.

\item[$\bullet$] {\em Scattering and Rescattering:} PYTHIA chooses partons to 
participate in the
hard scattering from each nucleon. The partons that are chosen, as well as the
momentum fraction they carry, are based on known parton distributions
\cite{cteq}. After the individual partons have had a hard scattering and are 
color-connected with the diquarks from the remaining nucleon, strings are 
formed. The kinematics of the fragments from the string are determined by 
JETSET and are stored in a temporary array. Any partonic radiation that 
is not color-connected to either string goes directly into the 
nucleus-nucleus final state.  This string is then put back into the nuclei
and allowed to rescatter as a ``wounded'' nucleon. The wounded nucleon 
has the string's momentum while its position is updated to halfway between 
the original nucleons' positions through, 

\narrowtext
\begin{equation}
(x_1,y_1) (x_2,y_2) \to (\frac{x_1+x_2}{2},\frac{y_1+y_2}{2}).
\end{equation}

\item[$\bullet$] {\em Final State:} After all nucleons have
experienced their geometrically determined number of collisions or they
have center of mass energies below the cutoff, particles present in 
each nucleon's temporary array constitute
the final state.

\end{itemize}

\narrowtext
Hadronic observables from the model have been compared against data for several
systems. Although the model is based on very simple premises, it reproduces the
main features characterizing hadronic final states. Most notable for
the present work, we have made comparisons to S+Au hadronic data as well 
as dilepton CERN SPS data (see section~\ref{sec:results} for dilepton data). 
Our model matches the $\pi^0$ spectrum from WA80 \cite{wa80}, as well 
as the 
$\pi^-$s
from NA35 \cite{na35}.  The two comparisons are shown in Figs. 1 and 2,
respectively. The $p_T$ distribution of neutral pions is 
slightly above the
data at low $p_T$ and slightly below the data for high $p_T$ pions.
This might account for part of the difference in kinematic 
acceptance discussed later in Sec~\ref{sec:norm}.

\section{Secondary Scattering}
\label{sec:secondary}

Dileptons from pseudoscalars ($\pi^0$, $\eta$, $\eta^{\prime}$) and vectors
($\omega$, $\rho^0$, $\phi$) produced in the primary
scattering phase are not enough to account for the S+Au data. 
Our model also incorporates secondary scattering of hadronic resonances. All
$\pi\,$s and $\rho\,$s formed during the primary collisions of nucleons will 
have a chance to scatter amongst themselves before
decaying. The reactions we consider are of two types, one which produces a
resonance that decays to dileptons and the other which goes to dileptons
directly.  

Of the first type, $\pi^+ \pi^- \to \rho^0$ $\to e^+
e^-$, $\pi^0 \rho^{\pm} \to {a_1}^{\pm}$ $\to \pi^{\pm} 
e^+ e^-$,
$\pi^{\pm} \rho^0 \to {a_1}^{\pm}$ $\to \pi^{\pm} e^+ e^-$ 
have been included. To accomplish these types of scattering, pions and 
rhos must of course appear in the final state of the model described in the
previous section. As the default, JETSET automatically decays all hadronic 
resonances, but it also contains provisions to prohibit them.
We thus allows neutral pions to scatter from charged rhos when
conditions are favorable.  Technically, the steps involved in secondary 
scattering are similar to those for primary scattering. 

\begin{itemize}

\item[$\bullet$] {\em Number of collisions:} The number of collisions is again
determined geometrically using the appropriate density and cross section. 

\item[$\bullet$] {\em Picking scattering partners:} A $\pi^+ \pi^-$ or $\pi
\rho$ pair is randomly 
chosen and allowed to scatter if (1) the pair has not already scattered, 
(2) the pair is moving toward one another in the transverse plane and
(3) the pair is within one cross-sectional radius of one
another in the transverse beam direction.
The cross section for creating a resonance is taken to
be

\narrowtext
\begin{equation}
\sigma(\sqrt{s})=\frac{\pi}{{\bbox{k}^2}}
\frac{{\Gamma_{\rm partial}}^2}{(\sqrt{s}-m_{\rm res})^2
+{\Gamma_{\rm full}}^2/4}
\end{equation} 

with $\bbox{k}$ being the center-of-mass momentum. The full and partial 
decay widths for $\rho^{0} \to \pi^{+} \pi^{-}$ are set to 152 MeV. The 
full $a_1$ 
decay width is 400 MeV and the partial width for ${a_1}^{\pm} \to 
\pi \rho$ are energy dependent\cite{xiong}:

\narrowtext
\begin{equation}
\Gamma_{a_1 \to \pi \rho}=\frac{{g_{a_1\pi\rho}}^2}{24 \pi {m_{a_1}}^2}
|{\bf{k}}| [2(p_{\pi} \cdot p_{\rho})^2+{m_{\rho}}^2({m_{\pi}}^2+{\bf{k}}^2)]
\label{eq:ga1pr}
\end{equation}
 
\item[$\bullet$] {\em Resonance formation and decay:} The kinematics of the
resonances are determined from the pair of hadrons while JETSET decays the 
resonance
into dileptons using appropriate functions for $d\Gamma/dM^2$ and 
${|\cal M|}^2$ resulting from analysis of the same
Lagrangian used to derive Eq. (\ref{eq:ga1pr}).

\end{itemize}

Dileptons from secondary scatterings of the resonance
type increase the number significantly in the region around the $\rho^0$ mass,
but not in the region with the largest gap, 0.2 GeV $\leq$ $M^2$ $\leq$ 0.5 
GeV.

The nonresonant component is estimated here by computing the sole process 
$\pi^0 \rho^{\pm} \to \pi^{\pm} e^+ e^-$ and then assuming the
other isospin channels contribute equally.  The other
channels involve Feynman graphs that result in a singularity and
must be regulated in a full $T$-matrix or some other effective 
approach \cite{khcg}.  To this level of estimate, isospin averaging and
ignoring interference effects between these and the resonant $a_{1}$ 
contributions is not worrisome. 
The prescription for directly scattering pions and rhos is very similar 
to the one used for resonance scattering as detailed below. 

\begin{itemize}
\item[$\bullet$] {\em Picking scattering partners:} 
The cross section for $\pi^0
\rho^{\pm} \to \pi^{\pm} \rho^0$ determines how many and how
often the charged rhos scatter with pions.  Using an effective Lagrangian
approach to be described momentarily, we generate
a set of graphs.  The resultant cross section can
be nicely parameterized for $\sqrt{s} \le$  2.5 GeV by

\narrowtext
\begin{equation}
\sigma(\sqrt{s})=1.8 \, {\rm mb}+{8\over s^{2}}\,{\rm mb}\, {\rm GeV}^4
+0.5\,s\, 
\left(\frac{\rm mb}{{\rm GeV}^2}\right)
\end{equation} 

\item[$\bullet$] {\em Scattering:} Since this is a non-resonant process,
the Monte Carlo directly determines the kinematics of the final state.
Necessary ingredients for such procedures include an interaction Lagrangian
and a resulting squared matrix element.
The Lagrangian employed is\cite{kapusta} 

\narrowtext
\begin{equation}
{\cal L}=|D_{\mu} \Phi|^{2}-{m_{\pi}}^{2} | \Phi |^{2}-{\frac{1}{4}} 
\rho_{\mu \nu} 
\rho^{\mu \nu}+{\frac{1}{2}} {m_{\rho}}^{2} \rho_{\mu} \rho^{\mu}-{\frac{1}{4}} 
F_{\mu \nu} F^{\mu \nu}
\end{equation}

where $D_{\mu} = \partial_{\mu}-ieA_{\mu}-ig_{\rho}{\rho}_{\mu}$ is the
covariant derivative, $\Phi$ is the complex charged pion field, 
${\rho}_{\mu \nu}$ is the rho
field-strength tensor and $F_{\mu \nu}$ is the photon field strength tensor.
From this Lagrangian, the matrix elements can be determined. In the
calculation, the graphs involving the $a_1$ are neglected as the contribution 
from $a_1$ has already taken into account in the resonance 
portion of the model. There are three graphs whose matrix
elements are listed below.

\narrowtext
\begin{equation}
{\cal M}_1=\frac{g_{\rho} e^2}{M^2 (s-{m_{\pi}}^2)} \epsilon^{\mu}(p_a)
(2p_b+p_a)_{\mu} (2p_1+q)_{\nu} \bar{u}(p_{-}) {\gamma}^{\nu} \bar{v}(p_{+})
\end{equation}

\narrowtext
\begin{equation}
{\cal M}_2=\frac{-h_{+}(t) g_{\rho} e^2}{M^2 (t-{m_{\rho}}^2)} 
\epsilon^{\mu}(p_a)
(2p_a-q)_{\nu} (p_b+p_1)_{\mu} \bar{u}(p_{-}) {\gamma}^{\nu} \bar{v}(p_{+})
\end{equation}

\narrowtext
\begin{equation}
{\cal M}_3=\frac{g_{\rho} e^2}{M^2} \epsilon^{\mu}(p_a) [X_{\mu \nu}]
\bar{u}(p_{-}) {\gamma}^{\nu} \bar{v}(p_{+})
\end{equation}

where $X_{\mu \nu} = a g_{\mu \nu} + b ({p_1}_{\mu} {p_b}_{\nu} + 
{p_b}_{\mu} {p_1}_{\nu}) + c ({p_b}_{\mu} {p_b}_{\nu} + 
{p_1}_{\mu} {p_1}_{\nu})$.

In the $t$-channel matrix element, a form 
factor, $h_{+}(t) = ({\Lambda^2 - m^2})/({\Lambda^2 - t})$, appears
to  account for the finite size of the mesons. Its presence 
breaks gauge invariance. In order to completely restore
gauge invariance, other terms must be added to the four point diagram 
${\cal M}_3$: $a=-1$ $b=c=({h_{+}(t)-1})/({{p_{b}} \cdot q 
+ {p_{1}} \cdot q})$. The parameters $\Lambda$ and $m$ are set to 1 GeV 
and $m_{\rho}$, respectively.

The absolute square $|{\cal M}_{1}+{\cal M}_{2}+{\cal M}_{3}|^2$ 
and $d\sigma/dM^2$ were used to Monte Carlo the three-body $\pi^{\pm} 
e^{-} e^{+}$ final state. 

\end{itemize}

These lepton pairs are of nonresonant origin and are now added to the pairs 
from resonance decays. 

Since the pions and rhos are scattering inside the reaction zone, their
dynamics are altered by the medium. Being of bremsstrahlung type, 
these mechanisms are therefore susceptible to the Landau-Pomeranchuk-Migdal 
effect\cite{lpa}.  Pions and rhos involved in secondary scattering will
undergo frequent multiple scatterings, and not only with other pions
and rhos. Therefore, the number of dileptons produced by this scattering is 
reduced. The reduction factor is dependent at minimum on the
invariant mass of the lepton pair as well as the mean free path of 
the pions and rhos.  We use a reduction 
$1-e^{-M \lambda}$, where $M$ is the invariant mass of the lepton pair
and $\lambda$ is the mean free path of the hadrons. For our purposes and 
level of estimation here, we set $\lambda$ to some average value $\sim 1$ 
fm\cite{hp}.

The total dilepton yield from our model is the sum of lepton pairs from primary
plus secondary scattering. The invariant mass distributions of 
the dileptons from all contributions will be discussed in the last section.

\section{Normalization}
\label{sec:norm}

In order to keep computation time low, the code was run to look at
dileptons from $\pi^0$, $\eta$, $\eta^{\prime}$, $\omega$, $\rho^0$, $\phi$
separately. In each run, the hadron considered was allowed to decay only 
into the dilepton channel, all other modes were prohibited. 
Technically, this is merely a way to maximize statistics.
To reinstate absolute normalization, all lepton
pairs counted were multiplied by the branching ratio for the process from
which they came. This approximation is valid because the lepton decay mode is a
comparatively rare event. We have successfully used this perturbative technique 
before to calculate high-energy photon production in the framework of BUU 
transport theory\cite{buu}. The secondary scattering resonance production 
was handled much the same way. All resonances produced from secondary 
scattering decayed exclusively into their lepton channels and were later 
multiplied by the appropriate branching ratio. Normalization 
procedures for non-resonance $\pi \rho$ scattering is somewhat different. 
Instead of a branching ratio from the
particle data book as before\cite{pdg}, the fraction of
dilepton events is based on the calculated cross section for $\pi \rho
\to \pi e^+ e^-$ divided by the total cross section for $\pi \rho$
scattering. For simplicity this total cross section is the sum of the elastic
cross section, $\sigma_{\pi^0 \rho^{\pm} \to \pi^{\pm} \rho^0}$ plus
the resonance cross section, $\sigma_{\pi^0 \rho^{\pm} \to {a_1}^{\pm}}$
(incoherently).

\section{Results}
\label{sec:results}

The invariant mass spectra of dileptons from the primary scattering part of
our model for three different systems are displayed in Fig. 3. The top two
plots display lepton pairs from p+Be collisions and p+Au collisions,
respectively. The lower plot shows dileptons in question from S+Au
collisions. The simulation agrees with the
proton-induced data and it is reassuring that our S+Au model-results are 
consistent with the cocktail from the CERES collaboration\cite{ceres}.
Plotting against the actual S+Au data reveals a significant enhancement over
predictions in the invariant mass region between 200 and 500 MeV. There is also
a modest enhancement for masses above this range.

Dileptons from
secondary scattering for the S+Au system in our model are shown in Fig. 4. 
The contribution from pion annihilation increases the distribution 
significantly
in the rho mass region, but still leaves an excess below the rho mass.
We should stress that we have taken vacuum rho properties throughout. 
Radiative $a_{1}$ decay contributes a minimal amount in 
the excess (or deficit) region, but the contribution from non-resonance 
$\pi \rho$ scattering provides the most significant increase. 

With the inclusion of the secondary scattering previously described, the 
invariant
mass distributions of dileptons are shown in Fig. 5. The dilepton spectra from
the proton-induced interactions are not significantly changed.
This result is as expected---dileptons from the
smaller systems are quantitatively described by primary hadronic decays. The 
proton-nucleus collisions do not create a heated nuclear medium large enough 
or dense enough to bring about significant collective effects. 
Conversely, the S+Au
collision has a marked increase in lepton-pair production between an invariant
mass of 200 and 500 MeV as well as a noticeable increase in the higher mass
region. It is not surprising that secondary scattering becomes important in the
S+Au system, as a dense nuclear medium is created during the collision.

Allowing for the short-comings of our model, results still suggest that
secondary scattering is a viable explanation for the excess found in dilepton
data. Inclusion of secondary scattering, 1) preserves the consistency the 
primary scattering in our model has with proton-induced data and, 2) enhances 
the number of
dileptons within the region of excess discovered in S+Au data. Although this
agreement cannot rule out other possible explanations for the excess 
electrons, our model's simplicity is attractive.

\begin{figure}
\caption{$\pi^0$ $p_T$ distribution from WA80 as compared
with the model.}
\label{wa80}
\end{figure}

\begin{figure}
\caption{$\pi^-$ rapidity distribution from NA35 compared
to the model.}
\label{NA35}
\end{figure}

\begin{figure}
\caption{Dilepton invariant mass spectra from primary scattering in
the model compared to CERES data.}
\label{ceres5}
\end{figure}

\begin{figure}
\caption{Contributions from secondary scattering in S+Au collision.}
\label{ceres2}
\end{figure}

\begin{figure}
\caption{Total dilepton invariant mass distributions, including primary and
secondary scattering in the model as compared with CERES data.}
\label{ceres6}
\end{figure}

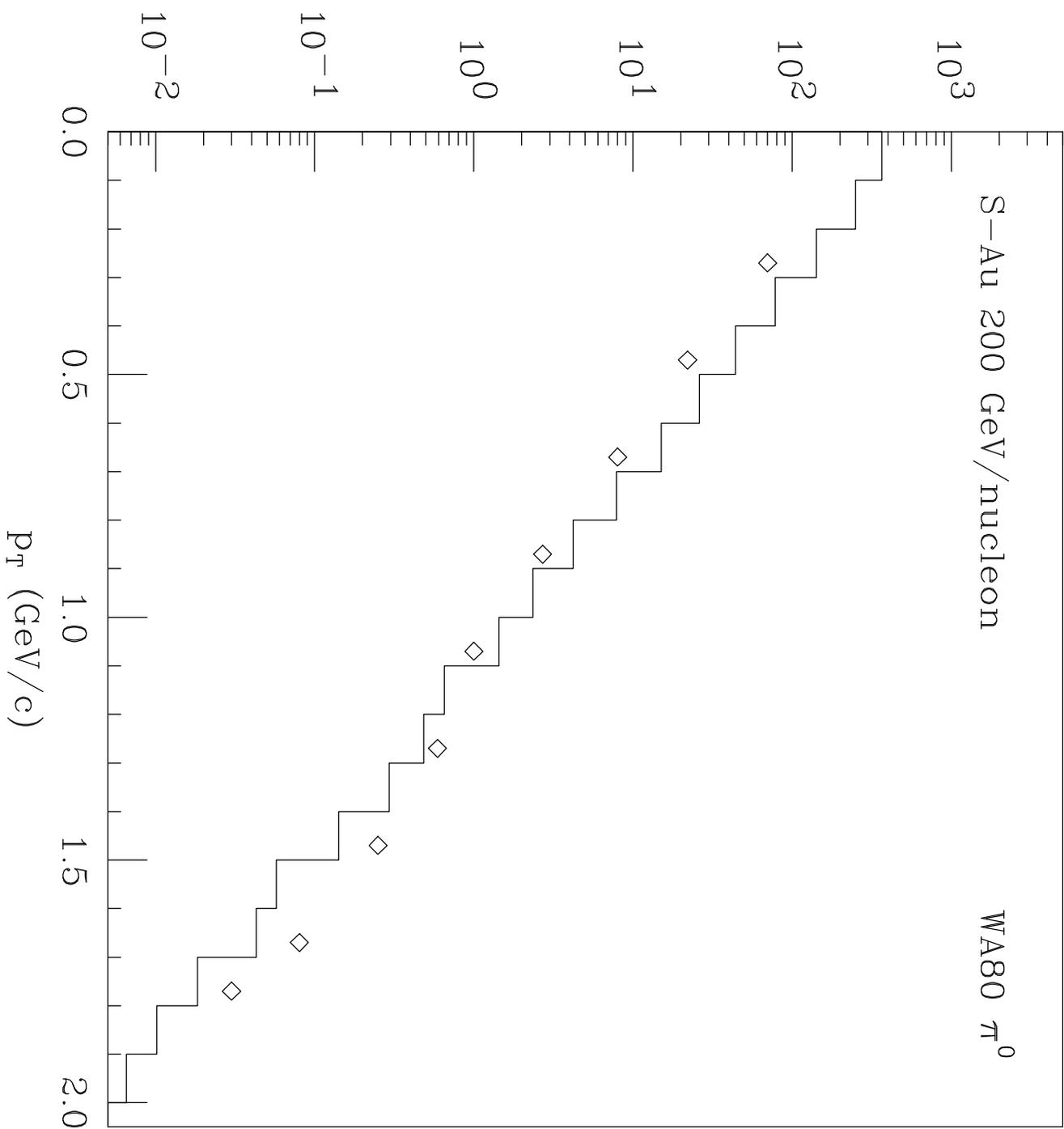
\end{document}